\documentclass{aa}
\usepackage{natbib}
\usepackage{epsfig}
\usepackage{csquotes}
\usepackage{tikz}
\usetikzlibrary{decorations.pathmorphing}
\usetikzlibrary{decorations.pathmorphing,calc}
\def\ut#1{\mathop{\vtop{\ialign{##\crcr
     $\hfil\displaystyle{#1}\hfil$\crcr\noalign
     {\kern1pt\nointerlineskip}\hbox{$\hfil\sim\hfil$}\crcr
     \noalign{\kern1pt}}}}}

\def\undersymbol#1#2{\mathop{\vtop{\ialign{##\crcr
     $\hfil\displaystyle{#2}\hfil$\crcr\noalign
     {\kern1pt\nointerlineskip}\hbox{$\hfil#1\hfil$}\crcr
     \noalign{\kern1pt}}}}}

\usepackage{graphicx}

\begin{document}

\title{Cosmic voids and the kinetic analysis\\ IV. Hubble tension and the cosmological constant}
       \author{V.G.Gurzadyan\inst{1,2}, N.N.Fimin\inst{3}, V.M.Chechetkin\inst{3,4} }

              \institute{Center for Cosmology and Astrophysics, Alikhanian National
Laboratory, Alikhanian Brothers str.2, 0036, and Yerevan State University, Manukian str.1, 0025 Yerevan, Armenia \and
SIA, Sapienza Universita di Roma, Via Salaria 851/881, 00191 Rome, Italy \and Keldysh Institute of Applied Mathematics of RAS, Miusskaya Sq. 4, 125047 Moscow, Russia \and Institute of Computer Aided Design of RAS, 2nd Brestskaya st., 123056 Moscow, Russia}

   \offprints{V.G. Gurzadyan, \email{gurzadyan@yerphi.am}}
   \date{Submitted: XXX; Accepted: XXX}

 \abstract{The formation of cosmic structures in the late Universe was considered using the Vlasov kinetic approach.  The crucial point is the use of the gravitational potential with repulsive term of the cosmological constant, which provides a solution to the Hubble tension, that is, the Hubble parameter for the late Universe has to differ from its global cosmological value.  This also provides a mechanism of formation of stationary semi-periodic gravitating structures of voids and walls, so that the cosmological constant has the role of the scaling. It can therefore be compared with the observational data for given regions. The considered mechanism of the structure formation in the late cosmological epoch then succeeds the epoch described by the evolution of primordial density fluctuations.
}


   \authorrunning{V.G. Gurzadyan, N.N.Fimin, V.M.Chechetkin}
   \titlerunning{Cosmic voids and the kinetic analysis. IV.}
   \maketitle
%

\section{Introduction}

The Hubble tension \citep{R,R1,R2,Val,Dai2,Scol1,Scol2} has attracted attention to the possibly genuine differences in the descriptions of the late and early Universe, including of the structure formation and the large-scale flows (e.g. \citep{photon,Bo,Baj,Cap1}). The problems of the nature of dark matter and dark energy are still unsolved, and diverse approaches and models are developed to consider the entire scope of the observational challenges (see \citep{Cap,Ped} and references therein).   

The evolution of primordial density fluctuations as the origin of the cosmic structure formation \citep{Peeb} at large scales is described by the hydrodynamical pancake theory of Zeldovich \citep{Z,SZ}. It predicts the formation of the cosmic web and the filaments. At smaller scales, scales of clusters and groups of galaxies, the role of local gravitational interaction dominates. Namely, while the hydrodynamic approach assumes little influence of the intrinsic gravity of the involved particles as compared to the influence on the geometry of inhomogeneities  2D caustics; \citealt{Arn,ArnP}) of the Friedmann flow, the kinetic field approach is assumed to be suitable for relatively fast non-equilibrium phenomena in an external quasi-ordered medium, when the wave fronts overturn
and in the intersection of three-dimensional matter flows channels for kinetic processes are formed \citep{GFC1,GFC2,GFC3}.

The possible tension of the late (local) and early (global) Universe outlines these differences in the mechanisms of structure formation and of the flows at large and small scales. Namely, the explanation of the Hubble tension as a result of two flows, a local and a global flow, with a non-identical Hubble constant, was provided in \citep{GS7,GS8} based on the theorem proved in \citep{G1}. This theorem states the general function that fulfils the equivalency of gravity of a sphere and of a point mass in the following form for the force 
\begin{equation}
F=-{\frac{GMm}{r^{2}}}+{\frac{\Lambda c^{2}mr}{3}}.
\end{equation}

This function ensures a non-force-free gravitational field inside a shell, and the second term in the left-hand side involves the cosmological constant in weak-field General Relativity and McCrea-Milne non-relativistic cosmological model \citep{MM,Z81}. Importantly, \citep{GS2,G2,GS3,GS4} showed that Eq.(1) enables us to describe the observational data of the dynamics of groups and clusters of galaxies. Moreover, observational data support the non-force-free field inside a shell, as predicted by Eq.(1), that is, the influence of the galactic haloes on the properties of the spiral galaxies \citep{Kr}. 

Then, the late Hubble flow is described by the equation \citep{GS7,GS8}
\begin{equation}\label{Hl}
        H_0^2 = \frac{8 \pi G \rho_0}{3} + \frac{\Lambda c^2}{3},
\end{equation}  
where $H_0$ is the local Hubble parameter, and $\rho_0$ is the local mean density of matter. This equation follows from Eq. (1) and has the structure of the standard Friedmann equation, but it has a different content within the non-relativistic McCrea-Milne model with a cosmological constant.
 
Below, we continue the Vlasov kinetic analysis \citep{GFC1,GFC2,GFC3} of the structure formation in the small-scale Universe. The essential point in the analysis is the role of the cosmological constant, and hence, that the evolutionary paths of the filament formation are the result of the self-consistent gravitational interaction of the particles along with the repulsive cosmological term. Namely, it appears that zones with a dominating influence of the ascending and descending branches of the modified gravitational potential of Eq.(1) form a topologically different
multi-connected  matter structure in the Universe. We analysed the of quasi-static processes that occurred in the system defined by the Vlasov-Poisson equations near its state of relative equilibrium. In this case, as shown, it is reasonable to reduce the problem to the analysis of the integral equation of a Hammerstein type. For the latter, the corresponding boundary problem is again based on the theorem in \citep{G1} because the structure of the
of solutions of this integral equation and of the possible singularities can be expressed using known formalisable representations. We investigated the properties of the spectrum of the non-linear integral operator and its relation to the spectrum of the Fredholm operator for the linear potential, which enabled us to reveal the deterministic way of the formation of coherent filaments in local regions in the Universe.

\section{The kinetic approach}

The starting point for obtaining hydrodynamic models with singularities, that is, caustics and kinetic models without taking into account the influence of gravity,
is the introduction of the Friedmann flow. This is characterized by a field of velocities whose values for each pair of points in space
are proportional to the distance between these points, ${\bf v}=H \Delta{\bf r}$.
The Friedmann flow in these models was considered as some analogue of an equilibrium state, which is perturbed either by inhomogeneity in the initial data set or by random fluctuations in the density
that lead to the formation of a geometrical small inhomogeneity (adiabatically stable) of the spatial distribution of particles of the system on the background of this flow. The smallness is caused by random deviations, which are assumed a priori to be essentially limited in norm in comparison with the basis values of the mean parameters.

The system of Vlasov--Poisson equations for describing of dynamics
in a system of $N$ cosmological objects (with masses $m_{i=1,...,N} =m\equiv 1$) may be represented as (see \citep{GFC1} for details)
$$
\frac {\partial F({\bf x},{\bf v},t)}{\partial t} +
{\rm {div}}_{\bf x}({\bf v}F)+\widehat {G}(F;F)=0,~~~
\widehat{G}(F; F) \equiv
$$ 
\begin{equation}
-\big ({\rm{\nabla}}_{\bf v}F\big )
\big({\nabla }_{\bf x}(\Phi [F({\bf x})]\big),
\label{1}
\end{equation}
\begin{equation}
{\Delta}_{\bf x}^{(3)}\Phi[F({\bf x})]\big|_{t=t_0}=
A S_3 G \int F( {\bf x},{\bf v},t_0)\:d{\bf v} - \frac{c^2\Lambda}{2},
\label{2}
\end{equation}
$$
S_3\equiv{\rm{meas} \:{\mathcal S}^2} =4\pi,~~{\mathcal S}^d=\{{\bf x}\in  \mathbb {R}^d, |{\bf x}=1|\},
$$
where $F({\bf x},{\bf v},t)$ is the distribution function of gravitationally interacting
particles,
$A$ is the normalization factor for the particle density, $t_{0}$ is a fixed
moment in time, and $G$ is the gravitational constant.
The system of objects or particles is considered  within the finite domain of configurational space
$\Omega \subseteq {\mathbb R}^3$ (${\rm {diam}\:\Omega}\le \infty$),
with a $C^2$-smooth boundary $\partial \Omega$.

Equation (\ref{2}) is the non-linear Poisson equation, which accounts for the cosmological term of Eq.(1).
The third term on the right-hand side of the kinetic equation (\ref{1}) may be represented as
\begin{equation}
\widehat{G}(F; F) =  {\bf G}(F)\frac{\partial F}{\partial {\bf v}},~~~
\label{3}
\end{equation}
$$
{\bf G}(F)=-\nabla_{\bf x}\Phi[F({\bf x})],
$$
$$
\Phi[F({\bf x})]= AS_3 G \int\int  {\mathfrak K}_3
({\bf x}-{\bf x}')  F({\bf x}',{\bf v}',t_*)\:d{\bf x}'d{\bf v}' +
$$
$$
\frac{\Lambda c^2}{12}|{\bf x}|^2 + \widehat{\mathfrak B}_3 ({\bf x},{\bf x}'),
$$
\noindent
where ${\mathfrak K}_3 ({\bf x}-{\bf x}')=-{|{\bf x}-{\bf x}'|^{-1}}$,
$\widehat{\mathfrak B}_3 ({\bf x},{\bf x}')$
is an operator term that takes into account the influence of the boundary conditions.
The classical  Newtonian potential
$\Phi_{N}(r)= -G m/r$ increases monotonically
in the interval $r\in (0,+ \infty )$ ($\Phi_N \in (-\infty,0)$), while the
generalised (with cosmological term) Newton gravity potential
$$
\Phi_{GN}(r)\equiv -G m/ r - \frac{1}{2}c^2\Lambda r^2, 
$$
has a maximum 
$$
\Phi_{GN}^{(max)} (r_c)= -\frac{1}{2}G(3mc^{2/3})\Lambda^{1/3},
$$
where $r_c = \big(G m/(3\Lambda c^2)\big)^{1/3}$ (it increases in the interval $r \in(0;\:r_c]$ and decreases in the interval  $r \in(r_c;\:\infty)$).

We consider the stationary case of dynamics,
$F= F( {\bf x},{\bf v})$. However, the further analysis
mainly concerns
the second equation of the system, which is the Poisson equation
relative to the potential, and no explicit time dependence is observed in it.
When it varies, the Hilbert--Einstein--Maxwell action
 \citep{Ved2} can therefore be a separate variation
over the fields (for a fixed particle distribution) and a variation
over the distribution functions
(with fixed fields). The approach we considered
is therefore applicable
for adiabatic processes at a quasi-equilibrium (weakly varying)
particle distribution functions.
In this case, we can use the energy
substitution for a unique variable of the distribution function \cite{Ved}: $F({\bf x},{\bf v})=f(\varepsilon)
\in C^1 _{+} ({\mathbb R}^1)$, where
$\varepsilon = m{\bf v}^2/2 + \Phi ({\bf x})$.
Thus, the particle density in the right side of the Poisson equation
can be expressed in terms of the equilibrium solution of the Vlasov equations.
This solution is identical to Maxwell--Boltzmann distributions $f=f_0(\varepsilon )=
AN\exp(-\varepsilon /\theta)$. However, the physical meaning of the equilibrium solution of the Vlasov equation
is essentially different from that of the Boltzmann equation. This solution must meet the following requirements:
1)\:a maximum possible statistical independence, 2)\:an isotropy of the velocity distribution, and 3)\:stationarity of 
the distribution in the form $F({\bf x},{\bf v})=\rho({\bf x})\prod_{i=1,2,3}{\mathfrak f}(v_i^2)$.
The substitution  expression into the Vlasov equation gives
\begin{equation}
\sum_i\bigg( v_i\frac{\partial \ln(\rho)}{\partial x_i}  - \frac{\partial \Phi}{m\partial x_i}  \frac{\partial {\mathfrak f}(v_i^2)}{{\mathfrak f}(v_i^2) \:\partial v_i}    \bigg)F=0,
\label{4}
\end{equation}
and we obtain a system of ODEs,
\begin{equation}
\frac{\partial(\ln\:\rho)/\partial x_i}{-\partial \Phi/\partial x_i} = \frac{\partial \ln\big({\mathfrak f}(v_i^2)\big/\partial v_i)}{m v_i}=  -\theta^{-1},
\label{5}
\end{equation}
\noindent
where $\theta$ is a constant of separation of variables, and its physical meaning is kinetic temperature in the
system of interacting collisionless particles (in accordance with Vlasov's definition \citep{V1,V2} ,
thermodynamic/collisional equilibrium is globally absent in this system).

Equation (\ref{2}) for the gravitational potential  can be written as
$$
\Delta \Phi({\bf x}) = ANG S_3^2
\bigg (\int_{y\in [0,\infty ]}
\exp \big(-y^2/(2\theta) \big)y^2\:dy
\bigg )\cdot 
$$
\begin{equation}
\exp(-\Phi/\theta)-\frac{c^2\Lambda}{2},~~~
A,\theta,R_\Omega \in {\mathbb R}^1,
\label{6}
\end{equation}
\noindent
where $R_\Omega$ is the radius of the region $\Omega$ and is accepted
in the form of a ball in configurational $3$--space (as the simplest physically realizable case).

The Poisson equation (\ref{2}) takes the form of the inhomogeneous equation of
Liouville--Gelfand \citep[LG;][]{Du}, in which the local (generalized) temperature changes sign
depending on the value of the derivative of the potential at a given point:
As mentioned above, for the two-particle problem (in particular, for a formal pair in the form of a centre
coalescence of the main part of the particles and the conditional extremely distant particle)
the repulsive force can dominate due to
the presence of a quadratic term $\sim |{\bf x}|^2$.
Generalised to the indefinite thermodynamics of a system of gravitating particles, this becomes
similar to that for the Onsager vortices in classical hydrodynamics \citep{FM}, and the existence of solutions of the LG equation for large system sizes ensures the existence of solutions to the Vlasov equation (\ref{1}).
This can be shown using the parametric Young inequality
\citep{Pokh}. It was shown \citep{GFC1} that for the
conditions $c^2\Lambda\gtrless 3\pi \lambda^\dag$ ($\theta \gtrless 0$), the system of Vlasov--Poisson equations
has solutions of the type of distribution functions that admit the energy
substitution, and the potential of
the gravitational field, which has the property of convexity (in the general case,
for an arbitrary $R_\Omega \le \infty$, in contrast to the case of the attraction potential, for
which there is a limit
$R_\Omega < \big(C_0 \theta^2 /  \big(\lambda^\dag/S_3^2 \big)^2  \big)^{1/4}$). We used the notation $\lambda^\dag\equiv ANG S_3^2 {\mathcal J}(\theta)$, ${\mathcal J}(\theta)\equiv
\int_0^{v_{max}} \exp\big(-v^2/(2\theta)\big)v^2dv$.

As already noted in \citep{GFC1},
in the formulation of the Dirichlet problem for the Poisson equation (\ref{2}) (or (\ref{6})) with a constant right-hand side
on the boundary of the $\Omega$ region according to a
McCrea-Milne averaging gravitational field
outside the compact subdomain $\Omega_0$ that contains a
system of particles (that is situated in
the region $\Omega$ (${\rm meas }\:{\Omega_0}\ll {\rm meas }\,{\Omega}$)), we can
assume the  boundary condition on the $\partial \Omega$  as given in accordance with the theorem in \citep{G1}.

The solution of the Dirichlet problem may be obtained with the help of
an integral representation of the equation for
a gravitational (double--layer) potential \citep{Nowakowski}. The equation for the potential with
a Maxwell--Boltzmann particle density, corresponding to
an internal Dirichlet problem in a bounded domain $\Omega$
(under boundary conditions corresponding to the Milne--McCrea model) has the following form:
\begin{equation}
\Phi ({\bf x}) = \lambda_I \int_{\Omega'}{\mathcal K}({\bf x},{\bf x}')
 \exp \big( -\Phi ({\bf x}')/\theta \big) d{\bf x}' - \frac{c^2\Lambda}{12}{\bf x}^2 + C_0',~~
\label{7}
\end{equation}
$$
{\mathcal G}({\bf x},{\bf x}') \equiv 4\pi \sum^\infty_{\ell =0}
\sum_{m=-\ell}^\ell \frac{Y_{\ell m}^{*} (\vartheta',\varphi')
Y_{\ell m} (\vartheta,\varphi)}{2\ell + 1}
\frac{x_{<}^\ell x_{>}^\ell}{{{R}}^{2\ell + 1}_\Omega},
$$
$$
x_{<}={\rm{min}}(|{\bf x}|,|{\bf x}'|),~~x_{>}={\rm{max}}(| {\bf x}|,|{\bf x}'|),
$$
$$
{\mathcal K}(|{\bf x}-{\bf x}'|)\equiv {\mathcal G}({\bf x},{\bf x}') -
\frac{1}{|{\bf x}-{\bf x}'|},~~~
C_0' = -\frac{G Nm}{{{R}_\Omega}} - \frac{c^2\Lambda {{R}_\Omega}^2}{12},
$$
$$
\lambda_I =\lambda^\dag /S_3.
$$
\noindent
In essence, the above is the explicit form of the equation for the potential introduced in
expression (\ref{3}), where the Green function ${\mathcal G}( {\bf x},{\bf x}')$
for the inner boundary value problem in the domain $\Omega$ (in this case, due to
symmetry of the latter, we have $\int_{\Omega'} {\mathcal G}( {\bf x},{\bf x}')\rho (|{\bf x}'|)d{\bf x}'\to
C_1={\rm const} (\propto{1/R_\Omega})$).
We introduce a new variable $U( {\bf x})\equiv (\Phi ({\bf x})-C_0'-C_1)/\theta +  \alpha|{\bf x}|^2$, $\alpha \equiv c^2\Lambda /(12\theta)$,
and the above equation can be written as the uniform Hammerstein integral equation,
\begin{equation}
U({\bf x})=
\lambda_\theta  \widehat {\mathfrak G}({U}),~~~
\widehat{\mathfrak G}({U})\equiv \int_{\Omega'} {\mathcal K}(|{\bf x}-{\bf y}|)\Psi\big({\bf y},U({\bf y})\big)\:d{\bf y},
\label{8}
\end{equation}
$$
\lambda_\theta \equiv \frac{\lambda^\dag}{\theta S_3}\exp\big((-C_0-C_1)/\theta\big),~~~
$$
$$
{\mathcal K}(|{\bf x}-{\bf y}|)=-|{\bf x}-{\bf y}|^{-1},~~~\Psi\big({\bf y},U({\bf y})\big)\equiv -\exp\big(-\alpha  {\bf y}^2 -U({\bf y}) \big).
$$
For $\theta >0$, we obtain $\lambda_\theta >0$,
the mapping $\widehat {\mathfrak G}({U})$ is a compact (in $L_2(\Omega)$) non-linear operator, since the conditions of the Nemytskij--Vainberg theorem \citep{Kras} are satisfied for it,
($\int_\Omega \int_\Omega {\mathcal K}^2(|{\bf x}-{\bf y}|)\:d{\bf x}d{\bf y}={\mathcal K}_\Omega <\infty$, $\Psi\big({\bf y},U({\bf y})\big)\in C(\Omega \otimes {\mathbb R})$ and
$|\Psi\big({\bf y},U({\bf y})\big)| \le g({\bf y})+ C_\Psi \cdot|{U}|$, $g \in L_2(\Omega), \:g({\bf y}),C_\Psi >0$).

In the expression $\Psi^\dag \equiv \int^{U}_0 \Psi\big({\bf y},U\big)dU=\exp(-\alpha  {\bf y}^2)\cdot(\exp(-U)-1)$ it is clear that
$\Psi^\dag \le \tau_1 |U| + \tau_2$ (${\bf x}\in \Omega$), where $\tau_{1,2}>0$, $\tau_{1}<1/\lambda_{\mathcal K}$, $\lambda_{\mathcal K}$ is the maximum eigenvalue of the integral 
equation kernel ${{\mathcal K}}$.
Then, in accordance with theorem 2.8 in \citep{Zabr}, the Hammerstein equation has at least one solution $U_0({\bf x})$. We consider the question 
of the uniqueness of a solution or the presence of many solutions in the next paragraph.

\section{Solutions and cosmological consequences}

We assumed that for $\lambda=(\lambda_{\theta})_0$ Eq. (\ref{8}) has a non-trivial solution $U=U_0$.
We considered the Fredholm determinant ${\mathcal D}({\lambda})$ \citep{Zem} for an integral kernel of the linearised (in the vicinity $O(U_0)$ of the basic solution $U_0$) Hammerstein equation
$\widetilde{\mathcal K}({\bf x},{\bf y})\equiv{\mathcal K}(|{\bf x}-{\bf y}|)\cdot\partial\Psi\big({\bf y},U_0({\bf y})\big)/\partial U_0({\bf y})$. 
After linearisation (an application of the Frechet derivative), we obtained the linear Fredholm self-adjoint compact operator with a discrete spectrum (on the real axis; \citealt{Kras}).
If ${\mathcal D}\big((\lambda_{\theta})_0\big)\neq 0$ (i.\:e. $(\lambda_{\theta})_0$ is not the characteristic value of the kernel $\widetilde{\mathcal K}({\bf x},{\bf y})$), 
then in the vicinity $O\big((\lambda_\theta)_0 \big)$
Eq. (\ref{8}) has a unique analytic (by powers of $\big(\lambda_\theta-(\lambda_\theta)_0 \big)^j$, $j=1,2,...$) solution $U({\bf x}|\lambda_\theta)$ \citep{Cord}, for 
which $\lim_{\lambda_\theta\to(\lambda_\theta)_0} U({\bf x}|\lambda_\theta)=U_0({\bf x})$ ($\forall {\bf x}\in \Omega$).

We considered the solution of (\ref{8}) in the vicinity $O(U_0)\times O(\lambda_\theta)_0$, and introduce the perturbed characteristic value  $\lambda_\theta = (\lambda_\theta)_0+\xi$ and
the perturbed solution $U=U_0({\bf x})+\zeta({\bf x})$. Following the method in \citep{Akhm}, we substituted these expressions in Eq. (\ref{8}), written in the  form 
$$
-\zeta ({\bf x})= (\lambda_\theta)_0 \int_\Omega  \frac{\big( 
\omega_1 ({\bf y})\zeta + \omega_2 ({\bf y})\zeta^2+...\big)d{\bf y}}{|{\bf x}-{\bf y}|} + 
$$
\begin{equation}
\xi \int_\Omega  \frac{\big(\omega_0 +\omega_1({\bf y})\zeta + \omega_2({\bf y})\zeta^2 +...  \big)d{\bf y}}{|{\bf x}-{\bf y}|},
\label{81}
\end{equation}\noindent
and taking into account the
Taylor expansion $\Psi\big({\bf x},U\big)=\sum_{j=0,1,...}\omega_j({\bf x})\zeta^j({\bf x})$ (here $\zeta ({\bf x})=\sum_{k=1}\xi^k \zeta_k ({\bf x})$),
$$
\omega_0 ({\bf x})=\Psi\big({\bf x},U_0\big)=-\exp\big(-\alpha  {\bf x}^2 -U_0({\bf x}) \big),
$$
\begin{equation}
\omega_1 ({\bf x})=\frac{\partial \Psi\big({\bf x},U\big)}{1!\partial U}\bigg|_{U=U_0}=\exp\big(-\alpha  {\bf x}^2 -U_0({\bf x}) \big),...,
\label{9}
\end{equation}
\noindent
we obtained a system of recurrent  linear and multi-linear equations for the variables $\zeta_i ({\bf x})$
$$
-\zeta_1 ({\bf x}) = (\lambda_\theta)_0 \int_\Omega  |{\bf x}-{\bf y}|^{-1} \omega_1({\bf y})\zeta_1 ({\bf y})\:d{\bf y}+
$$
\begin{equation}
(\lambda_\theta)_0\int_\Omega  |{\bf x}-{\bf y}|^{-1} \omega_0({\bf y})\:d{\bf y},
\label{10}
\end{equation}
$$
-\zeta_2 ({\bf x}) = (\lambda_\theta)_0 \int_\Omega  |{\bf x}-{\bf y}|^{-1} \omega_1({\bf y})\zeta_2 ({\bf y})\:d{\bf y}+
$$
\begin{equation}
\int_\Omega  |{\bf x}-{\bf y}|^{-1} \big( (\lambda_\theta)_0 \omega_2({\bf y}) \zeta_1^2({\bf y}) + \omega_1({\bf y})  \zeta_1({\bf y}) \big)\:d{\bf y},...
\label{11}
\end{equation}
Since, by assumption, ${\mathcal D}({\lambda}_0)\neq 0$, then for a linear non-uniform Fredholm $II$ type equation  (\ref{10}) there exists a resolvent $R({\bf x},{\bf y};\lambda_0)$, and
we can write
$$
\zeta_k ({\bf x}) =-(\lambda_\theta)_0 \int_\Omega  R({\bf x},{\bf y};\lambda_0) {\mathcal H}_k \big(\zeta_1({\bf y}),\zeta_1({\bf y}),...,
$$
\begin{equation}
\zeta_{k-1}({\bf y})\big)\:d{\bf y}+ 
{\mathcal H}_k \big(\zeta_1({\bf x}),\zeta_1({\bf x}),...,\zeta_{k-1}({\bf x})\big)
\label{12}
\end{equation}
\noindent
(here the operator function ${\mathcal H}_k$ is the sum of all integrals including $\zeta_1,\zeta_2,...$ up to $(k-1)$--th power: 
$\zeta_k=-(\lambda_\theta)_0\int |{\bf x}-{\bf y}|^{-1}\zeta_1({\bf y})d{\bf y}+{\mathcal H}_k({\bf x})$).

Thus, we consistently and uniquely defined all functions $\zeta_k$. Consequently, we constructed a power series for 
$\zeta ({\bf x})=\zeta \big(\xi^i,\zeta_i({\bf x})\big)\equiv \xi^i\zeta_i$ (summation over repeating indices), and
it only remains to prove that the constructed series
converges (for all values from some convergence circle) for the deviation parameter $\xi$ (this completely establishes the equivalence of the expansion
$\zeta \big(\xi^i,\zeta_i({\bf x})\big)$ and the function  $\zeta ({\bf x})$).

Based on the properties of the Lyapunov--Schmidt integral operator with a weakly polar kernel (lemma 8.1 in \citep{Kras}), and explicit forms of 
$\omega_k \propto \exp(-\alpha {\bf x}^2-U)/k!$,
we can state $|\int_\Omega |{\bf x}-{\bf y}|^{-1}\omega_k ({\bf y})d{\bf y}|<Z_1(={\rm const})$ ($k=0,1,...$, $\forall {\bf x}\in \Omega$). Next, by definition, 
$|R({\bf x},{\bf y};\lambda_0)|<Z_2 (={\rm const})$ ($\forall {\bf x},{\bf y}\in \Omega$). Then, we can  write the major series for series $\zeta \big(\xi^i,\zeta_i({\bf x})\big)$.
We introduced an algebraic equation $\zeta^\dag =Z_3 \big( \xi + \xi \zeta^\dag + (\xi + ({\lambda_\theta})_0)\cdot\big((\zeta^\dag)^2 + (\zeta^\dag)^3 +...\big)   \big)$, 
$Z_3 \equiv Z_1(1+Z_2 |({\lambda_\theta})_0|)$.
We substituted $\zeta ({\bf x})= \xi^j \varkappa_j|_{j=1,2,...}$ into the last equation and then compared the coefficients at different powers of the deviation variable $\xi$,
\begin{equation}
\varkappa_1 = Z_3,~~~\varkappa_2 = Z_3 (\varkappa_1 + |({\lambda_\theta})_0|\cdot\varkappa_1^2),
\label{13}
\end{equation}
$$
\varkappa_3 = Z_3 \big(   \varkappa_2 +\varkappa_1^2 + 2  |({\lambda_\theta})_0|\cdot\varkappa_1  \varkappa_2 + |({\lambda_\theta})_0| \varkappa_1^3 \big),... 
$$
\noindent
Consequently, $|\zeta_k({\bf x})|<\varkappa_k$ (${\bf x}\in \Omega$), and the convergence region of the series $\zeta =Z_3 \big( \xi + \xi \zeta + ...  \big)$ is equivalent to the 
convergence region  of the series $\zeta ({\bf x})=\xi^j \zeta_j({\bf x})|_{j=1,2,...}$.  To define the convergence region of $\zeta=\xi^i\varkappa_i$, we ought to investigate
the implicit function  $\zeta =Z_3 \big( \xi + \xi \zeta + ...  \big)$ as a function of $\zeta=\widetilde{\zeta}(\xi)$
\begin{equation}
\widetilde{\zeta}(\xi) = Z_3\xi + (Z_3 \xi -1)\zeta^\dag + Z_3 (\xi + |({\lambda_\theta})_0|)\cdot( (\zeta^\dag)^2 + (\zeta^\dag)^3+...)=0.
\label{14}
\end{equation}
By the implicit function theorem \citep{Deim}, since $\partial \widetilde{\zeta}/\partial \zeta^\dag=-1$ for $\xi=\zeta^\dag=0$, then there exists a convergence 
circle with a positive radius for the series 
$\zeta^\dag = Z_3(\xi + \xi \zeta^\dag + (\xi + |({\lambda_\theta})_0|)(  (\zeta^\dag)^2 + (\zeta^\dag)^3+...)$. Consequently, there exists a function $\zeta ({\bf x})$, and the function
$$
U({\bf x})  =   U_0({\bf x})  +    \zeta({\bf x}) = U_0({\bf x})  + 
$$
\begin{equation}
\big( \lambda_\theta -  ({\lambda_\theta})_0 \big)\zeta_1 ({\bf x}) + 
\big( \lambda_\theta -  ({\lambda_\theta})_0 \big)^2 \zeta_2 ({\bf x}) + ...,
\label{15}
\end{equation}
\noindent
which is a holomorphic solution of the Hammerstein equation (\ref{8}) in the vicinity $O\big(({\lambda_\theta})_0 \big)$ (herewith $\lim_{({\lambda_\theta})\to ({\lambda_\theta})_0} U=U_0$).

We recall some of the properties of the linear compact operator $\widehat{L}\phi=\lambda\int_\Omega {\mathcal K}(|{\bf x}-{\bf y}|)\phi ({\bf y})d{\bf y}$. 
The kernel of this operator is symmetric, of weak polar type;
consequently, the operator $\widehat{L}$ belongs to the Hilbert--Schmidt type of operators.
The existence of at least one characteristic function has been known since Kellogg \citep{Kel}. Moreover, there exists a sequence of characteristic values and corresponding 
eigenfunctions of the investigated kernel of the linear integral operator (theorem 146 in \citep{Tit}).
In accordance with \citep{Newt}, we considered
\begin{equation}
\lambda_{\ell,j} = R_\Omega^{-2} \cdot\big(   \varphi_j^{(\ell +1/2)}    \big)^{2}, ~~~\ell\ge 0,~~j\ge 1,
\label{16}
\end{equation}
\noindent
where $\varphi_j^{(\ell +1/2)}$ are the roots of the  transcendental equation
$$
(2\ell + 1) J_{\ell + 1/2} \big( \varphi_j^{(\ell +1/2)}  \big) + \frac{\varphi_j^{(\ell +1/2)}}{2}\big( J_{\ell - 1/2} \big( \varphi_j^{(\ell +1/2)}  \big) -
$$ 
\begin{equation}
J_{\ell + 3/2}
\big( \varphi_j^{(\ell +1/2)}  \big) \big)=0,
\label{17}
\end{equation}
\noindent
where $J_\nu(...)|_{\nu \in {\mathbb R}}$ refers to the Bessel function of fractional order. The eigenfunctions corresponding to each eigenvalue
$\lambda_{\ell, j}$ can be represented in spherical coordinates in the form $\phi_{\ell, j,m}(r,\theta,\chi)=J_{\ell+1/2}(\sqrt{\lambda_{\ell,j}}r)Y^m_\ell(\theta,\chi)$ ($|m|\le \ell$),
$Y^m_\ell (\theta,\chi)=P^m_\ell \big({\rm{cos}}(\theta)\big)\cos(m\chi)$. It should be noted that we considered the Hammerstein equation (\ref{8}) for  an arbitrarily shaped region $\Omega$,
but for a spherical region,  we assumed $\ell=m=0$ ($Y^0_0 = \sqrt{1/(4\pi)}$). The kernel  
$\widetilde{\mathcal K}({\bf x},{\bf y})=-|{\bf x}-{\bf y}|^{-1}\omega_1 ({\bf y}, U_0(\bf y))$ of the Fredholm equation for 
the potential belongs to the Schmidt class, and it can be transformed into symmetrical form,
\begin{equation}
\widetilde{\mathcal K}({\bf x},{\bf y})= -\sqrt{\omega_1({\bf y})/\omega_1({\bf x})}|{\bf x}-{\bf y}|^{-1}\sqrt{\omega_1({\bf y}) \omega_1({\bf x})}.
\label{171}
\end{equation}
\noindent
In this case, the resolvent for kernel $\widetilde{\mathcal K}({\bf x},{\bf y})$ is equal to $-\sqrt{\omega_1({\bf y})/\omega_1({\bf x})}R_1({\bf x},{\bf y};\lambda)$,
where $R_1$ is a resolvent for the symmetric kernel $-|{\bf x}-{\bf y}|^{-1}\sqrt{\omega_1({\bf y}) \omega_1({\bf x})}$. In fact, $\omega_1({\bf y})$ is a weight for the Newtonian kernel.
 We denote $\lambda_1 \equiv ({\lambda_\theta})_0 = {\lambda}_{0,1} (\widetilde{\mathcal K})$ as the characteristic number to which the eigenfunction corresponds $\phi_1\sim J_{1/2}(\sqrt{\lambda_1}r)$. 
To simplify the further calculations, we assumed $\omega_1\equiv 1$ (this can always be accomplished by multiplying of the left- and right-hand sides of
 (\ref{81}) by $\sqrt{\omega_1}$ and redefining $\sqrt{\omega_1}\zeta \to \zeta'$, $\omega_0/\sqrt{\omega_1}\to\omega_0'$ etc.). We searched for solutions of
 (\ref{81}) in the form of a series $\zeta = \xi^i\zeta_i|_{i=1,2,...}$. After substituting the last series, we obtained a system of  recurrent integral equations,
 \begin{equation}
- \zeta_1 ({\bf x})= \lambda_1 \int_\Omega |{\bf x}-{\bf y}|^{-1}\zeta_1({\bf y})d{\bf y} + \int_{\Omega}|{\bf x}-{\bf y}|^{-1}\omega_0d{\bf y},
 \label{18}
\end{equation}
 $$
  -\zeta_2 ({\bf x})= \lambda_1 \int_\Omega |{\bf x}-{\bf y}|^{-1}\zeta_2({\bf y})d{\bf y} +  \int_{\Omega}|{\bf x}-{\bf y}|^{-1} \big( \lambda_1 \omega_2  
  \zeta_2^2({\bf y}) +
$$
$$      
\zeta_1({\bf y}) \big)d{\bf y},...
 $$
 According to the Fredholm alternative, the condition of the existence of the solution for the first equation of the system is the orthogonality of the characteristic function $\phi_1$ and of 
 the second term in the right-hand side of (19): $\widehat{\mathfrak{H}}(\phi_1)\equiv\int_{\Omega}\omega_0({\bf y})\phi_1({\bf y})d{\bf y}=0$. This case is physically unrealizable (for $\phi_1,\phi_2,...$),  
  which can be checked directly. This means the absence (in the neighborhood of the characteristic value $\lambda_k$, $k\ge 1$) of  the analytic solution of the non-linear  equation for
 the gravitational potential.
 Therefore, we turn to case $\widehat{\mathfrak{H}}(\phi_1)\neq 0$. Then, Eq. (\ref{11}) is unsolvable (the condition of the Fredholm theorem is absent),  and consequently, the analytic series
 $\zeta=\xi^i\zeta_i$ (for integer indices and powers) does not exist. However, we can consider the representation $\zeta({\bf x})$ as a Puiseux series:   
 $\zeta=\xi^{k/2}\zeta_k|_{k=1,2,...}$. We denote $\xi^{1/2}\equiv \nu$, and then, equation (\ref{81}) takes the form
 $$
- \zeta ({\bf x}) = {\nu}^2 \int_\Omega |{\bf x}-{\bf y}|^{-1}\big( \omega_0 + \zeta ({\bf y}) + \omega_2 \zeta^2 ({\bf y})  +...  \big)d{\bf y}+
$$
\begin{equation}
\lambda_1
 \int_\Omega  |{\bf x}-{\bf y}|^{-1}\big(  \zeta ({\bf y}) + \omega_2 \zeta^2 ({\bf y})  +...   \big)d{\bf y},
  \label{19}
\end{equation}
 and as mentioned above, the  Puiseux series takes the form $\zeta=\nu^k\zeta_k|_{k=1,2,...}$. Substituting this series into equation (\ref{19}), we obtain an infinite system of equations,
  \begin{equation}
-\zeta_1 ({\bf x})= \lambda_1 \int_\Omega |{\bf x}-{\bf y}|^{-1} \zeta_1 ({\bf y})\:d{\bf y},
   \label{20}
\end{equation}
$$
-\zeta_2 ({\bf x})= \lambda_1 \int_\Omega |{\bf x}-{\bf y}|^{-1} \zeta_2 ({\bf y})\:d{\bf y} +
$$
\begin{equation}
 \int_\Omega |{\bf x}-{\bf y}|^{-1} \big(  \lambda_1 \omega_2 \zeta_1^2 ({\bf y}) +  
\zeta_1 ({\bf y})   \big)d{\bf y},~...,
   \label{21}
\end{equation}
$$
-\zeta_n ({\bf x})= \lambda_1\int_\Omega |{\bf x}-{\bf y}|^{-1} \zeta_n ({\bf y})\:d{\bf y} +
$$
\begin{equation}
\int_\Omega |{\bf x}-{\bf y}|^{-1} \big(
2 \lambda_1  \omega_2 \zeta_1 ({\bf y})\zeta_{n-1} ({\bf y}) + {\mathfrak Q}(\zeta_1,...,\zeta_{n-1})
\big)d{\bf y}.~~~
   \label{22}
\end{equation}
Then, we obtain  $\zeta_1 ({\bf x})= {\mathfrak E}_1\cdot \phi_1({\bf x})$, ${\mathfrak E}_1={\rm const}.$ The condition (by the Fredholm alternative) for the existence of a solution of equation (\ref{21}) can be written as 
\begin{equation}
\int_\Omega\int_\Omega |{\bf x}-{\bf y}|^{-1} \big(   
\lambda_1 \omega_2 \zeta_1^2 ({\bf y}) + \omega_0
\big) \phi_1({\bf x})d{{\bf x}}d{{\bf y}}=0,
\label{23}
\end{equation}
or, after integration by the variable ${\bf x}$: $\int\big(\lambda_1 \omega_2 \zeta_1^2 ({\bf y}) + \omega_0\big) \phi_1({\bf y})d{{\bf y}}=0$. When we substitute
the obtained above expression $\zeta_1 = {\mathfrak E}_1\cdot \phi_1$ in this formula, then we have for the constant ${\mathfrak E}_1$ an explicit expression,
\begin{equation}
{\mathfrak E}_1 = \pm \sqrt{-\frac{\int_\Omega \omega_0 \phi_1 ({\bf y})d{\bf y}}{\int_\Omega \lambda_1 \omega_2 \zeta_1^3 ({\bf y}) d{\bf y} }}
\label{24}.
\end{equation}
Equation (\ref{21}) may be written as
$$
-\zeta_2 ({\bf x}) = {\mathcal P}_2 ({\bf x}) + {\mathfrak E}_1\phi_1 ({\bf x}),~~  {\mathcal P}_2 ({\bf x})
$$
\begin{equation}
\equiv \sum_{j=2,...} \frac{\phi_j ({\bf x})}{\lambda_j-\lambda_1}\int_\Omega
\big(   
\lambda_1 \omega_2 \zeta_1^2 ({\bf y}) + \omega_0
\big)\phi_j({\bf y})d{\bf y}.
\label{25}
\end{equation}
In the general case $n\ge 3$, we obtain
\begin{equation}
-\zeta_3 ({\bf x}) = {\mathcal P}_2 ({\bf x})  + 
{\mathfrak E}_3\phi_1 ({\bf x}),~~   
\label{26}
\end{equation}
$$
{\mathcal P}_2 ({\bf x})\equiv  \sum_{j=2,...} \frac{\phi_j}{\lambda_j-\lambda_1}\int_\Omega    
\big(2  \lambda_1 \omega_2 \zeta_1 ({\bf y}) \zeta_2 ({\bf y}) +
$$
$$
{\mathfrak Q}(\zeta_1,\zeta_{2})\big)\phi_j ({\bf y})d{\bf y},
$$
\begin{equation}
-\zeta_n ({\bf x}) = {\mathcal P}_n ({\bf x}) 
   +
{\mathfrak E}_n\phi_1 ({\bf x}),
\label{27}
\end{equation}
$$
{\mathcal P}_n ({\bf x}) \equiv \sum_{j=2,...} \frac{\phi_j}{\lambda_j-\lambda_1}\int_\Omega   
 \big(2  \lambda_1 \omega_2 \zeta_1 ({\bf y}) \zeta_{n-1} ({\bf y}) +
$$
$$
{\mathfrak Q}(\zeta_1,\zeta_{2},...,\zeta_{n-1})\big)\phi_j ({\bf y})d{\bf y},
$$
$$
\frac{{\mathfrak E}_n}{{\mathfrak E}_1}\int_\Omega \omega_0 \phi_1 ({\bf y})d{\bf y}= \int_\Omega \big(2\lambda_1 \omega_2 \zeta_1 ({\bf y}){\mathcal P}_n ({\bf y}) +
$$
$$
{\mathfrak Q}(\zeta_1,\zeta_{2},...,\zeta_{n+1})\big)\phi_1 ({\bf y}) d{\bf y}.
$$
Here, ${\mathcal P}_n ({\bf x})$ can be associated with the Ruston pseudo-resolvent \citep{Ruston}.
Consequently, we can write the series $\zeta=\xi^{k/2}\zeta_k|_{k=1,2,...}$, and this series  formally satisfies Eq. (\ref{81}). We ought to demonstrate the
convergence of this series in the neighbourhood $O(\xi)$. We introduce 1)\:the constant ${\mathfrak X}_0$, defined by the condition 
$|{\mathfrak E}_1 \phi_1 ({\bf x})|= |{\zeta_1 ({\bf x})}|<{\mathfrak X}_0$ ($\forall {\bf x}\in \Omega$), and 
2)\:the function $S(z)= (|\lambda_1| + \nu^2){\mathfrak M}z^2/(1-z/\rho^\ddag)+\nu^2(\omega_0^{(m)} + z)$, where $|\omega_0|<\omega_0^{(m)}$, $\rho^\ddag \in (0,\:\rho^\ddag_{max})$, $\rho^\ddag_{max}$
is the convergence radius of the series $\omega_2({\bf x})+\omega_3 ({\bf x})z + \omega_4 ({\bf x})z^2 +...$, $|\omega_2({\bf x})|<{\mathfrak M}$. We substitute in the definition $S(z)$
 the decomposition $z=\nu {\mathfrak X}_0 + \nu^2 ({\mathfrak X}_1 + {\mathfrak Y}_1) + \nu^3 ({\mathfrak X}_1 + {\mathfrak Y}_1)+...$ and formally
 obtain the series $S(z) =\nu^2 S_2 + \nu^3 S_3 +...$. The value $S_n(z)$ is a major function for expression $2\lambda_1 \omega_2 \zeta_1({\bf x}) \zeta_{n-1}({\bf x}) +
  {\mathfrak Q}(\zeta_1,\zeta_{2},...,\zeta_{n})$, and the condition ${\mathfrak X}_n + {\mathfrak Y}_n$ is a majorant function for $\zeta_{n+1}({\bf x})$, $n=1,2,...$

We denote 1)\:$(1/{\mathfrak E}_1)\int_\Omega \omega_0\phi_1({\bf x})d{\bf x}={\mathfrak N}(={\rm const})$, 2)\:${\mathfrak Y}_n=S_{n+1} \rho^\ddag_{max}$, and
3)\:$({\mathfrak N} + 2 |\lambda_1| {\mathfrak M} {\mathfrak X}_0 a^2){\mathfrak X}_{n}= S_{n+1}a^2$ ($a>|\phi_1({\bf x})|$). Consequently,
$|{\mathcal P}_n ({\bf x})|<{\mathfrak Y}_n$, ${\mathfrak E}_{n+1}<{\mathfrak X}_n$.
We introduce the functions ${\mathfrak X} \equiv \nu^2{\mathfrak X}_1 +  \nu^3{\mathfrak X}_2 +...$, ${\mathfrak Y} \equiv \nu^2{\mathfrak Y}_1 +  \nu^3{\mathfrak Y}_2 +...$, and then, the 
above given definition of variables ${\mathfrak X}_1,...,{\mathfrak X}_n,...$ and ${\mathfrak Y}_1,...,{\mathfrak Y}_n,...$ is equivalent to solving the system of equations
$$
{\mathfrak Y} = \rho^\ddag_{max} \big(
(|\lambda_1|+\nu^2)\frac{{\mathfrak M} (\nu {\mathfrak X}_0 +{\mathfrak X} +{\mathfrak Y})^2}{1 -(\nu {\mathfrak X}_0 +{\mathfrak X} +{\mathfrak Y})/\rho^\ddag} + 
$$
\begin{equation}
\nu^2 (\omega_0^{(m)} + \nu {\mathfrak X}_0 +{\mathfrak Y} +{\mathfrak X})
\big),
\label{28}
\end{equation}
\begin{equation}
(|{\mathfrak N}|+2|\lambda_1| {\mathfrak M}  {\mathfrak X}_0a^2 )\nu {\mathfrak X}
= a^2 \bigg( (\nu^2 +|\lambda_1|)\frac{{\mathfrak M} (\nu {\mathfrak X}_0 +{\mathfrak X} +{\mathfrak Y})^2}{1- (\nu {\mathfrak X}_0 +{\mathfrak X} +{\mathfrak Y})/\rho^\ddag} +
\label{29}
\end{equation}
$$
+\nu^2 \big( \rho^\ddag_{max} +\nu  {\mathfrak X}_0 +{\mathfrak X} +{\mathfrak Y} \big)-\nu^2 (\rho^\ddag_{max}  + {\mathfrak M} |\lambda_1| {\mathfrak X}_0^2)
\bigg).
$$
We replace the variables ${\mathfrak X} =\nu {\mathfrak X}^\dag$,  ${\mathfrak Y} =\nu {\mathfrak Y}^\dag$.
Then, the system (\ref{28})--(\ref{29}) takes the form
$$
\Theta_1 = {\mathfrak Y}^\dag -\nu {\mathfrak Y}^\dag \frac{{\mathfrak X}_0 +{\mathfrak Y}^\dag +{\mathfrak X}^\dag}{\rho^\ddag} - \rho^\ddag_{max} 
\bigg( (\nu^2 + |\lambda_1|)
$$
\begin{equation}
{\mathfrak M} \nu ({\mathfrak X}_0 + {\mathfrak Y}^\dag +{\mathfrak X}^\dag)
+
\label{30}
\end{equation}
$$
+\nu \big(\rho^\ddag_{max}  +\nu ({\mathfrak X}_0 + {\mathfrak Y}^\dag +{\mathfrak X}^\dag) \big)
\big(  1-\nu \frac{{\mathfrak X}_0 + {\mathfrak Y}^\dag +{\mathfrak X}^\dag}{\rho^\ddag}        \big)
\bigg)=0,
$$
$$
\Theta_2 = (|{\mathfrak N}|  + 2|\lambda_1|{\mathfrak X}_0 {\mathfrak M} a^2){\mathfrak X}^\dag - (|{\mathfrak N}|  + 
$$
\begin{equation}
2|\lambda_1|{\mathfrak X}_0 {\mathfrak M} a^2){\mathfrak X}^\dag \nu
\frac{{\mathfrak X}_0 + {\mathfrak Y}^\dag +{\mathfrak X}^\dag}{\rho^\ddag}-
\label{31}
\end{equation}
$$
-(|\lambda_1| + \nu^2) {\mathfrak M}  ({\mathfrak X}_0 + {\mathfrak Y}^\dag +{\mathfrak X}^\dag)^2 a^2 - 
$$
$$
-\big( \nu  ({\mathfrak X}_0 + {\mathfrak Y}^\dag +{\mathfrak X}^\dag) - {\mathfrak M} |\lambda_1|  {\mathfrak X}_0^2 \big)
\big( 
1- \nu \frac{{\mathfrak X}_0 + {\mathfrak Y}^\dag +{\mathfrak X}^\dag}{\rho^\ddag}
\big)=0.
$$
The Jacobi determinant of the latter system of equations is $\Delta = D(\Theta_1,\Theta_2)/D({\mathfrak X}^\dag,{\mathfrak Y}^\dag) =-|{\mathfrak N}|<0$ for ${\mathfrak X}^\dag={\mathfrak Y}^\dag=\nu=0$.
Consequently, the series of defined variables  ${\mathfrak X}$ and  ${\mathfrak Y}$ converges in the vicinity of the point $\nu=0$.
The series $\nu {\mathfrak X}_0 + \nu^2 ({\mathfrak X}_1 + {\mathfrak Y}_1) + \nu^3 ({\mathfrak X}_2 + {\mathfrak Y}_2) +...$  has a convergence circle (with a centre in the point $\nu=0$).
This indicates that the Puiseux series  $\zeta=\xi^{k/2}\zeta_k|_{k=1,2,...}$ converges in the vicinity of the point $\xi=0$.

We can conclude that the Hammerstein equation for the gravitational potential in the vicinity $O(\lambda_1)\ni \lambda$ (where $\lambda_1$ is one of the characteristic values 
of the linear Fredholm equation for the Newton potential) has two non-holomorphic solutions of the form
$$
U({\bf x}) =U_0 ({\bf x}) + (\lambda_{\theta}-({\lambda_\theta})_0)^{1/2}\zeta_1({\bf x}) + 
$$
\begin{equation}
(\lambda_{\theta}-({\lambda_\theta})_0)\zeta_2({\bf x}) +
(\lambda_{\theta}-({\lambda_\theta})_0)^{3/2}\zeta_3({\bf x}) +...
\label{32}
\end{equation}
We return to Eq. (\ref{8}) and to the assumption that $\big(({\lambda_\theta})_0, U_0\big)$ is a characteristic value and the eigenvalue of  the Hammerstein operator 
$\widehat{\mathfrak G}({U})$.
The above analysis of the structure of the solutions of the Hammerstein equation used the fact that the kinetic temperature in the system was assumed to be positive ($\theta >0$).
 For negative temperatures ($\theta <0$), the properties of the solutions of the equation for the potential change slightly, although the method of their constructive continuation 
on the parameter plane remains the same. Since the presence of negative temperatures in the system is due to the effect of anti-gravity (due to the cosmological term in the integral equation), we assumed the Newtonian kernel to be continuous and a bounded function (although we can also use the Hilbert-Schmidt theory), and, on the basis of theorem 2 in \citep{Jingxian}, we conclude that Eq. (\ref{8}) has at least two analytic solutions outside the points of the spectrum of the Fréchet derivative of the Hammerstein operator 
$(\lambda_{\theta <0})_k (\widehat{\mathfrak G}')$. In the vicinity of the characteristic points corresponding to the solutions of the non-linear equation and simultaneous coincidence with the special points of 
the resolvent of the linearized equation, the solutions are continued algebraically and can be written out in the form of Puiseux series on half-integer degrees of the deviation from the solution points.

In physical terms, the expression (\ref{32}) can be written as
$$
\Phi ({\bf x})= (C^\ddag_1 + c^2\Lambda/12)|{\bf x}|^2 + U_0({\bf x}) \sum_i \big( ANG{\mathcal J(\theta)}\exp(C^\ddag_1/\theta)-
$$
\begin{equation}
 ({\lambda_\theta})_0 \big)^{i/2},
\label{321}
\end{equation}
$$
{\mathcal J(\theta >0)} = -\exp(-v_{max}^2/(2\theta))v_{max} \theta + 
$$
$$
\theta^{3/2}\sqrt{\pi/2}\:{\rm erf}\big( v_{max}/ \sqrt{2\theta}  \big),
$$
$$
{\mathcal J(\theta <0)} =  \exp(v_{max}^2/(2\theta))v_{max} \theta - 
$$
$$
\theta \sqrt{-\pi \theta/2}\:{\rm erf}\big( v_{max}/\sqrt{-2\theta} \big),
$$
\noindent
where $U_0({\bf x})$ is a solution of the Hammerstein equation for the potential (existing due to the properties of the corresponding integral operator), 
$\lambda_1 (={\lambda_\theta})_0)$ is the characteristic
value of kernel $\omega_1/|{\bf x}-{\bf y}|$ (for the Fredholm equation) in the vicinity of the point $\lambda_\theta \in (({\lambda_\theta})_0-\epsilon,({\lambda_\theta})_0+\epsilon)$.
If ${sgn}\big(\int_\Omega \omega_0 \phi_1({\bf y})d{\bf y} \big) \neq  {sgn}\big(\int_\Omega \lambda_1\omega_2 \phi_1^3({\bf y})d{\bf y} \big)$, then for $\lambda >\lambda_1$
there exist two solutions, and for $\lambda <\lambda_1$ there are no solutions. For 
${sgn}\big(\int_\Omega \omega_0 \phi_1({\bf y})d{\bf y} \big) = {sgn}\big(\int_\Omega \lambda_1\omega_2 \phi_1^3({\bf y})d{\bf y} \big)$, then for $\lambda <\lambda_1$
there exist two solutions, and for $\lambda >\lambda_1$ there are no solutions.

Thus, we showed that the solution of the system of Vlasov-Poisson equations in the case of using the energy substitution in the stationary Vlasov equation and choosing the 
quasi-Maxwell distribution (depending on the kinetic temperature) as a basis solution of the kinetic equation, leads to the non-linear
Hammerstein integral equation for the potential, including repulsion due to the cosmological term in Eq.(1). The Hammerstein equation in essence
is an equation for the self-consistent field in the system of massive many-particles on cosmological scales. 
This leads to the simultaneous realisation in the cosmological system with a self-consistent field of two types of ordering: A significant increase in the
matter density and its decrease (at the one-dimensional motion of matter in the channel, walls appear with a high value of the gravitational potential, and voids, which are practically 
empty spaces between walls). Oscillations of the eigenfunctions even in the simplest case of spherical symmetry
 of the Green function can form analogues of interference patterns, which leads to the production of structure formation in a homogeneous medium far from the line that connects the massive objects.
 Since the eigenfunctions of the linearised version of the equation for the potential are proportional to the Bessel function (with a rapidly decreasing
weight function), the loss of periodicity of the solutions in physical space becomes obvious. Since for non-linear integral operators it is possible to establish the
the continuum character of the spectrum, we can observe the effect of secondary solutions (on the hyperplane of parameters near the basic solution of the equation) that coexist with the initial solutions,
which leads to the construction of the second type of periodicity: The translation of the solution isotropically along all directions. Thus, the composition of solutions with a potential of two types
in the presence of additional translational transgression due to the structure of the spectrum results in a distribution of the field and matter in space that is peculiar to the cosmic web. The scaling of the semi-periodic structures corresponds to the scale of the involved constant, that is, the cosmological constant in agreement with observational data \citep{GFC2,GFC3}. Incidentally, the change in the kinetic temperature at the zero-point transition can lead to a dipole-type structures involving a repeller.

The study performed above of the properties of solutions of the non-linear inhomogeneous integral Hammerstein equation obtained from the Dirichlet boundary value problem allowed us without the need to invoke the assumption of a correlation of fluctuations to describe in a completely deterministic manner the existence in the vicinity of equilibrium points of many-particle systems (at different scales) defined by secondary solutions with significantly different properties. We also proved in this analysis that these solutions exist non-locally. The presence of analytical and non-holomorphic branches of solutions of the equation for the potential given by Eq.(10) from a physical point of view in distinguished directions with sharp gradients of the density parameter of the medium has to be associated with the walls as the internal spaces between the voids. It should be noted that the cosmological term in the Hammerstein equation leads to the necessity of introducing of the concept of a modified kinetic temperature as defined by Vlasov \citep{V1,V2}. In contrast to the linear analysis based on the properties of the Fredholm equations of the second kind, which, as was previously established by the authors \citep{GFC1,GFC2,GFC3}, should lead to periodic and mixed-periodic cosmological structures, the method proposed in this study is more refined and suitable for numerical modelling. It can be used in certain cases as an alternative to statistical computations.

\section{Conclusions}

We performed an analysis of the filament formation in the late Universe based on the Vlasov kinetic technique and using the modified law of gravitational interaction with a cosmological constant, Eq.(1). While the primordial density fluctuations were considered to lead to an early structure formation via the Zeldovich pancake theory, that is, via stochastic dynamics, the Vlasov self-consistent field mechanism can provide structure features at late stages of the evolution. 

We continued the rigorous analysis of the Vlasov-Poisson equations of \citep{GFC1,GFC2,GFC3}. The equations reduced to the Hammerstein integral equation were shown to predict semi-periodic structures, that is, those peculiar to the cosmic web. The role of the repulsive term,  that is, of the cosmological constant, is a key role because it is also defines the scaling of the formed semi-periodic structures, that is, of the voids. In this context, the considered mechanism of the formation of the structures in the local Universe complements the role of the cosmological constant term in Eq.(1) in the appearance of two galactic flows. The local flow is determined by the in McCrea-Milne non-relativistic model,  and the global flow is determined by the Friedmann equations, while the discrepancy of the Hubble constants of the flows thus naturally explains the Hubble tension. 

Thus, the Hubble tension \citep{R1,R2,Scol1,Scol2} and other observational indications of the possible genuine difference in the properties of the early and late Universe can serve as tests for the mechanism of the structure formation on the local scale, as discussed above. In particular, the measurement by the Dark Energy Spectroscopic Instrument (DESI) collaboration \citep{Said} of the local value of the Hubble constant
and the distance to the Coma cluster using the fundamental plane of galaxies confirmed this tension. An additional informative window for tracing the cosmological evolution arises from the released DESI first-year data \citep{DESI1,DESI2,DESI3} on the
baryon acoustic oscillation using several tracers. The data revealed additional tensions with the $\Lambda CDM$ model. These observational data
support the approach of our study, that is, the need of developing different, more refined descriptions of the properties of the late Universe. In particular, this approach, in which the scaling radius $r_c = \big(G m/(3\Lambda c^2)\big)^{1/3}$ defined the role of the cosmological constant term in Eq.(1), enabled us to fit the dynamics of the galactic flow in the vicinity of the Virgo supercluster or the Laniakea supercluster, for instance \citep{GS7,GS8}.

The revelation of the dominating role of the self-consistent gravitational field in the formation of the cosmic structures on local cosmological scales can be considered the main result of our rigorous analysis. The refined comparison of the predictions of the considered mechanism with the observational surveys on the voids and of the filaments, for example, to study the distributions of their scales versus redshift, can therefore be of particular interest.

\begin{acknowledgements} 
We are thankful to the referee for valuable comments. VMC is acknowledging the Russian Science Foundation grant 20-11-20165. 
\end{acknowledgements}

\end{document}